# CMB Maps at 0°.5 Resolution
# I: Full-Sky Simulations and Basic Results

G. Hinshaw[1,2], C.L. Bennett[3], and A. Kogut[1]

## ABSTRACT

We have simulated full-sky maps of the Cosmic Microwave Background (CMB) anisotropy expected from Cold Dark Matter (CDM) models at 0°.5 and 1°.0 angular resolution. Statistical properties of the maps are presented as a function of sky coverage, angular resolution, and instrument noise, and the implications of these results for observability of the Doppler peak are discussed. The *rms* fluctuations in a map are not a particularly robust probe of the existence of a Doppler peak, however, a full correlation analysis can provide reasonable sensitivity. We find that sensitivity to the Doppler peak depends primarily on the fraction of sky covered, and only secondarily on the angular resolution and noise level. Color plates and one-dimensional scans of the maps are presented to visually illustrate the anisotropies.

## 1. Introduction

Observations of the cosmic microwave background anisotropy are fundamental to understanding the formation and evolution of structure in the universe. On angular scales larger than a few degrees, the CMB anisotropy traces the primordial density distribution, while measurements at higher resolution ($\theta \lesssim 1°$) probe the physical scales and causal mechanisms responsible for currently observed structure, offering greater power to distinguish between competing models of structure formation (Efstathiou, Bond, & White 1992; Coulson et al. 1994). Several groups have reported detections of anisotropy on degree angular scales (Schuster et al. 1993; Wollack et al. 1994; Cheng et al. 1994; De Bernardis et al. 1994; Dragovan et al. 1994;

[1]Hughes STX Corporation, Code 685, Laboratory for Astronomy and Solar Physics, NASA/GSFC, Greenbelt MD 20771.

[2]e-mail: hinshaw@stars.gsfc.nasa.gov

[3]Code 685, Laboratory for Astronomy and Solar Physics, NASA/GSFC, Greenbelt MD 20771.

Devlin et al. 1994; Clapp et al. 1994) but the implications of these experiments for models of structure formation are unclear.

Interpretation of observational results in a specific patch of sky are complicated by the fact that cosmological models predict the CMB anisotropy to be a single realization of a stochastic process whose properties (power spectrum, phase correlations, etc.) are predicted only for an ensemble average of equivalent realizations. Any single realization may be expected to vary about the ensemble average ("cosmic variance"). The generally small sky coverage of most existing observations (typically $10^{-3}$ to $10^{-4}$ of the full sky) provides another complicating factor: observations of small patches of the sky may not be representative of the sky as a whole, particularly in non-Gaussian models with distinctive rare features ("sample variance").

Although a large body of work exists providing analytic or Monte Carlo derivation of ensemble average properties of various models, relatively little has been published on the phenomenological properties expected for single medium-scale experiments. Questions of importance both to observers planning future instruments and to theorists analyzing existing data include the following: what is the role of sample variance in statistical analyses of small regions of the sky? What fraction of the sky must be observed to constrain models at specified confidence, particularly against Type II errors, accepting a hypothesis as true when it is false? What is the relation between limiting sources of uncertainty: the signal-to-noise ratio per pixel, cosmic variance, sample variance, and sky coverage? Are large-area scans with relatively greater instrument noise per pixel better than small-area deep observations? To what extent do features of the standard model such as the Doppler peak appear to observers as anomalies like isolated point sources? How important is it to produce a map of the microwave sky, as opposed to observing chopped scans?

Some of these questions are beginning to be addressed (Scott, Srednicki, & White 1994; Luo 1994; White 1994; Bond 1994) but many remain unanswered. In this Letter, we describe Monte Carlo simulations of CMB anisotropy in CDM models at $0.^\circ5$ and $1.^\circ0$ angular resolution, which we use to quantify the power of simple statistical tests that can be applied to the maps. As an example, we construct and evaluate a likelihood function for the universal baryon density, $\Omega_b$, and show that the sensitivity to $\Omega_b$ is determined primarily by the fraction of sky covered, and, to a lesser extent, by the angular resolution, and the instrument sensitivity. Further results on the analysis of unresolved features are reported in a companion paper (Kogut et al. 1995).

## 2. Map Simulations

We have simulated full-sky CMB maps at $0.^\circ5$ and $1.^\circ0$ angular resolution, Gaussian full width at half-maximum (FWHM), for a range of cosmologically interesting CDM parameters. The anisotropies expected in such models are expressed in terms of the mean angular power spectrum, specified by multipole amplitudes $C_\ell$, which, for Gaussian fluctuations, completely determine the enesemble-average properties of the sky. Figure 1, after Stompor (1994), shows the power spectra of the CDM models we consider; in all cases we adopt a scale-invariant primordial spectrum of fluctuations ($n = 1$) with a quadrupole normalization of $Q_{rms-PS} = 20\ \mu$K based on the two-year *COBE*-DMR data (Bennett et al. 1994; Górski et al. 1994; Wright et al. 1994; Banday et al. 1994). The dashed lines in the Figure represent the spectra convolved with $0\overset{\circ}{.}5$ and $1\overset{\circ}{.}0$ beams. A familiar feature in these spectra is the existence of the so-called Doppler peak at $\ell \simeq 220$. For a given primordial Sachs-Wolfe spectrum and a given Hubble constant, the amplitude of the first Doppler maximum increases with the baryon density $\Omega_b$. In this Letter, we restrict attention to flat models with $\Omega_0 = 1$, but it should be noted that the location of the first Doppler maximum, which probes the angular size of the horizon at the epoch of last scattering, depends primarily on $\Omega_0$: $\ell_{max} \sim 200/\Omega_0^{1/2}$ (Kamionkowski, Spergal & Sugiyama 1994). Thus, while a $1^\circ$ beam is adequate for probing the Doppler peak region in a flat universe (see §4.), it is marginal in a open universe.

To generate a particular realization of the CMB temperature observed at an instrumental dispersion $\sigma_b$ we evaluate the sum over spherical harmonics, $\Delta T(\theta,\phi) = \sum_{\ell,m}\ a_{\ell m} Y_{\ell m}(\theta,\phi) e^{-\frac{1}{2}\ell(\ell+1)\sigma_b^2}$, where the amplitudes $a_{\ell m}$ ($2 \leq \ell \leq 500$) are Gaussian random variables of zero mean and variance $C_\ell$, ($\langle a_{\ell m} a^*_{\ell' m'}\rangle = C_\ell\, \delta_{\ell\ell'}\, \delta_{mm'}$) appropriate to the model under consideration (Stompor 1994; Bond & Efstathiou 1987). Each map is pixelized according to the quadrilateralized spherical cube projection (White & Stemwedel 1992), which projects the entire sky onto six cube faces. Each face is divided into $2^{2(N-1)}$ approximately equal area, square pixels where $(N-1)$ is the index level of the pixelization. This construction allows for very efficient analysis of, for example, subsets of high resolution pixels contained within a given low resolution pixel. The current maps are pixelized with $N = 9$, which gives 393216 pixels of size $\sim 0\overset{\circ}{.}32 \times 0\overset{\circ}{.}32$.

Figure 2 shows color images of selected maps at $0\overset{\circ}{.}5$ angular resolution to illustrate visual features of the anisotropies expected in CDM models. Each map was generated with an independent random number seed, so the large scale features vary from map to map, but in all cases the structure due to the lowest order multipoles ($\ell \lesssim 10$) is readily apparent, and statistically equivalent. More striking differences are apparent on smaller angular scales where the power spectrum exhibits significant structure. The upper map was generated with a flat spectrum, defined as $C_\ell/4\pi = 1.2\, Q^2_{rms-PS}/(\ell(\ell+1))$, while the middle and lower maps were generated with $\Omega_b = 0.05, 0.20$ CDM spectra respectively. The increase in high frequency power evident from top to bottom is a plain manifestation of the increase in Doppler peak amplitude. In fact, the high frequency power in these maps is, in some cases, visually remniscent of a population of unresolved sources in the microwave sky. We explore this subject in a companion paper (Kogut et al. 1995). The peak amplitudes in these $0\overset{\circ}{.}5$ resolution maps typically range from $\sim 200\ \mu$K in the flat spectrum model to $\sim 400\ \mu$K in the $\Omega_b = 0.20$ CDM model.

Most present medium-scale experiments present their data in the form of one-dimensional, differential scans of a strip of sky. We crudely mimic such scans in Figure 3 where the left-hand panels show selected map profiles along a $10^\circ$ strip of sky near the north pole, and the right-hand panels show the corresponding single difference observations using a beam separation of $1\overset{\circ}{.}28$. (We have not subtracted a best fit gradient from the difference scans, as many observers do.) Note that relatively large and seemingly isolated features are not uncommon in these profiles.

## 3. *rms* Analysis

A basic measure of anisotropy in the maps is the pixel-pixel *rms* temperature fluctuation. Table 1 gives the *rms* fluctuations observed in the simulated maps as a function of the baryon content of the universe, the solid angle of the region selected for observation, and the angular resolution of the observations. The first row in each table gives the expected *rms*, and its standard deviation, for a full-sky map generated from the corresponding model parameters. The second row lists the *rms* observed in each individual full-sky map. In no case does the deviation from the ensemble average exceed $2\sigma$. The third row lists the mean *rms* observed in six patches of the sky, each subtending $\frac{1}{6}$th of the sky, as defined by the six pixelization cube faces. The uncertainty in this row is the standard deviation of the six observations and is given to illustrate the typical level of sample variance that can be expected in such observations. Successive rows in the table list the mean *rms* fluctuations observed in successively smaller (and more numerous) portions of the sky. In particular, the final row presents results for the 1536 patches of size $\sim 5^\circ \times 5^\circ$ ($8.2 \times 10^{-3}$ sr), which is comparable to, or larger than, the coverage in most current medium-scale observations.

The trend towards decreasing *rms* values with decreasing solid angle merely reflects the fact that small patches are insensitive to features whose angular scale is larger than the patch size. Such structure is subtracted off as part of the mean temperature in the patch. The increasing scatter with decreasing solid angle reflects the corresponding growth of sample variance. Note that the smallest patches considered here have a sample variance on the order of 15 – 20%, which is comparable to the cosmic variance in the full-sky *COBE* maps at low $\ell$. The trend towards increasing *rms* with increasing $\Omega_b$ is a result of the growth of the Doppler peak with $\Omega_b$. The scatter in the observed *rms* provides a crude means for estimating how well the height of the Doppler peak can be resolved for a given angular resolution and sky coverage, in the absence of experimental uncertainty. For example, a map covering $\frac{1}{6}$th of the sky with $1^\circ$ resolution (see Table 1) can resolve a Doppler peak characterized by $\Omega_b = 0.05$ from a model with no Doppler peak, but a map covering $20^\circ \times 20^\circ$ cannot, based on *rms* values alone. The situation is much better at $0\overset{\circ}{.}5$ resolution where, even in the smallest patches, the difference between an $\Omega_b = 0.05$ CDM sky and a flat spectrum sky is roughly $3\sigma$.

## 4. Likelihood Analysis

By definition, the *rms* fluctuations in a map are only sensitive to the pixel variance which, in turn, depends only on the total power within the window function of the experiment

$$\langle t_i t_i \rangle = \frac{1}{4\pi} \sum_\ell (2\ell + 1)\, B_\ell^2\, C_\ell$$

where $t_i$ is the temperature in pixel i of a map, the angled brackets denote a universal ensemble average, $B_\ell^2 = e^{-\ell(\ell+1)\sigma_b^2}$ is the beam factor and $C_\ell$ is the power spectrum. However, the angular power spectrum also predicts the correlations expected between pairs of pixels and, for Gaussian fluctuations, the covariance matrix

$$M_{ij} = \langle t_i t_j \rangle = \frac{1}{4\pi} \sum_\ell (2\ell + 1)\, B_\ell^2\, C_\ell\, P_l(\cos\alpha_{ij})$$

fully characterizes the map. Here $P_l(\cos\alpha_{ij})$ is the Legendre polynomial of order $\ell$, and $\alpha_{ij}$ is the angle between pixels $i$ and $j$. The likelihood of observing a map with pixel temperatures $\vec{t}$, given a power spectrum $C_l$, is

$$P(\vec{t})\, d\vec{t} = \frac{d\vec{t}}{(2\pi)^{N/2}} \frac{e^{-\frac{1}{2}\vec{t}^T \cdot M^{-1} \cdot \vec{t}}}{\sqrt{\det M}}$$

where $N$ is the number of pixels in the map. Assuming a uniform prior distribution of model parameters, the likelihood of a model, given a map $t$, is then

$$L(C_\ell) \propto \frac{e^{-\frac{1}{2}\vec{t}^T \cdot M^{-1}(C_\ell) \cdot \vec{t}}}{\sqrt{\det M(C_\ell)}}.$$

In principal, the additional information contained in the likelihood function should provide superior sensitivity to the such features as the height of the Doppler peak, as parameterized by $\Omega_b$. To test this, we evaluate the above likelihood as a function of $\Omega_b$ for the two smallest patch sizes considered above at 0°.5 and 1°.0 resolution. More precisely, we simulate 100 independent patches of size 10°.4 × 10°.4 and 5°.2 × 5°.2 at each resolution using the parameters $Q_{rms-PS} = 20\ \mu$K, $n = 1$, $H_0 = 50$ km s$^{-1}$ Mpc$^{-1}$, and $\Omega_b = 0.05$. Each 10°.4 (5°.2) patch contains 1024 (256) square pixels. We simulate uncorrelated receiver noise in the maps by adding a random Gaussian number to each pixel with a fixed standard deviation (this implies uniform sky coverage in the patch, we then modify the above covariance matrix to account for this contribution: $M_{ij} \rightarrow M_{ij} + \sigma\delta_{ij}$, where $\sigma$ is the noise standard deviation per pixel). According to Table 1 the *rms* fluctuations due to an $\Omega_b = 0.05$ model are typically $50 - 70\ \mu$K for these patches, so we add receiver noise ranging from $5 - 40\ \mu$K per pixel to probe a noise-to-signal regime from $\sim 0$ to $\sim 1$.

Figure 4 shows the likelihood functions obtained from single, randomly selected patches of the sky for each resolution and patch size. The width of the likelihood function, which measures the sensitivity to $\Omega_b$, is nearly independent of the angular resolution of the map (in the small range considered), but strongly sensitive to the fraction of sky observed. This point is reiterated in Table 2 which gives the statistics of the recovered maximum likelihood values for the ensemble of patches. In each case, the median of the 100 determinations is always within $2\sigma$ of the input, $\Omega_b = 0.05$, while the standard deviation ranges from 0.12 in the large, high-resolution, low-noise patches, to 0.41 in the small, high-noise, low-resolution patches. This table also demonstrates that in the regime of moderately high signal-to-noise, $\sigma \lesssim 40\ \mu$K, the sensitivity to $\Omega_b$ is nearly independent of noise level. We interpret these results as follows: a precise determination of the height of the Doppler peak requires a large number of "typical" high spatial frequency ($\ell \sim 200$) peaks and troughs to determine a "typical" amplitude. Moreover, given sufficient instrument sensitivity, a $1^\circ$ beam will not dilute these features to the point of non-detection. We have also evaluated the above likelihood function, in each of our simulated $\Omega_b = 0.05$ patches, for the flat spectrum model. In every patch the likelihood of this model is exponentially suppressed relative to the standard CDM models.

Of course the above analysis makes the optimistic assumption that all other cosmological parameters are known exactly and that there are no significant systematic effects (see, e.g., Wilkinson 1994) or foreground contaminants (see, e.g. Bennett et al. 1992). There is also considerable degeneracy amongst cosmological parameters with regard to the interpretation of the CMB anisotropy power spectrum in the sense that different parameter combinations can yield similar angular power spectra (Bond et al. 1994). The conclusions we draw about experimental sensitivities to $\Omega_b$ are subject to such considerations, however our conclusions regarding the experimental sensitivities to the CMB power spectrum itself are valid independent of theoretical models. Our general conclusion is that, given limited observing time, optimal sensitivity to the CMB power spectrum is obtained by covering as large a portion of the sky as possible consistent with obtaining a moderately high signal to noise ratio.

We are very grateful to Radek Stompor for providing us with CDM power spectra in advance of publication. Also, it is a pleasure to acknowledge useful conversations with Tony Banday and Krys Górski.

Table 1. pixel-pixel *rms* fluctuations ($\mu$K)

| Patch size[a] | | Model[b] | | |
|---|---|---|---|---|
| | flat[c] | $\Omega_b = 0.01$ | $\Omega_b = 0.05$ | $\Omega_b = 0.20$ |
| | | $0\overset{\circ}{.}5$ resolution | | |
| ensemble[d] | $66.7 \pm 2.5$ | $80.4 \pm 2.1$ | $85.7 \pm 2.0$ | $105.9 \pm 1.6$ |
| full sky | 62.3 | 82.5 | 88.1 | 105.9 |
| $\sim 90^\circ \times 90^\circ$ | $62.0 \pm 2.7$ | $81.4 \pm 5.7$ | $86.0 \pm 4.3$ | $105.7 \pm 2.3$ |
| $\sim 45^\circ \times 45^\circ$ | $60.2 \pm 5.7$ | $77.2 \pm 5.1$ | $82.2 \pm 4.1$ | $104.8 \pm 3.6$ |
| $\sim 20^\circ \times 20^\circ$ | $55.7 \pm 6.4$ | $72.8 \pm 5.6$ | $78.9 \pm 5.4$ | $100.7 \pm 4.8$ |
| $\sim 10^\circ \times 10^\circ$ | $49.9 \pm 7.3$ | $67.6 \pm 6.9$ | $73.9 \pm 6.5$ | $96.5 \pm 6.4$ |
| $\sim 5^\circ \times 5^\circ$ | $42.8 \pm 8.0$ | $61.4 \pm 8.2$ | $68.5 \pm 9.1$ | $92.1 \pm 10.1$ |
| | | $1\overset{\circ}{.}0$ resolution | | |
| ensemble | $61.5 \pm 2.7$ | $68.2 \pm 2.5$ | $71.2 \pm 2.3$ | $81.0 \pm 2.1$ |
| full sky | 58.8 | 70.3 | 72.5 | 80.7 |
| $\sim 90^\circ \times 90^\circ$ | $58.3 \pm 6.7$ | $68.5 \pm 5.0$ | $72.1 \pm 3.3$ | $78.7 \pm 2.3$ |
| $\sim 45^\circ \times 45^\circ$ | $55.8 \pm 7.0$ | $63.3 \pm 4.2$ | $66.8 \pm 6.0$ | $76.9 \pm 5.0$ |
| $\sim 20^\circ \times 20^\circ$ | $50.0 \pm 7.3$ | $58.3 \pm 6.9$ | $61.2 \pm 5.6$ | $73.5 \pm 6.3$ |
| $\sim 10^\circ \times 10^\circ$ | $43.1 \pm 7.9$ | $52.0 \pm 7.1$ | $55.4 \pm 7.3$ | $67.7 \pm 7.5$ |
| $\sim 5^\circ \times 5^\circ$ | $35.1 \pm 8.7$ | $44.6 \pm 8.5$ | $48.7 \pm 9.0$ | $61.3 \pm 10.7$ |

[a]The patch sizes listed above are approximate, for simplicity. The exact patch sizes, based on solid angles subtended by the low resolution pixels, are: $4\pi$, $4\pi/6$, $4\pi/24$, $4\pi/96$, $4\pi/384$, $4\pi/1536$.

[b]All other model parameters for this Table were fixed: $Q_{rms-PS} = 20$ $\mu$K, $n = 1$, and $H_0 = 50$ km s$^{-1}$ Mpc$^{-1}$.

[c]The flat model is defined as $C_\ell/4\pi = 1.2\, Q^2_{rms-PS}/(\ell(\ell+1))$.

[d]The ensemble average *rms* and its standard deviation expected from a given model.

Table 2. Derived $\Omega_b$ likelihood maxima from $\Omega_b = 0.05$ simulations[a]

| Patch size | Noise per pixel ($\mu$K)[b] | | | |
|---|---|---|---|---|
| | 5 | 10 | 20 | 40 |
| | | 0°5 resolution | | |
| 5°2 × 5°2 | $0.051 \pm 0.025$ | $0.051 \pm 0.027$ | $0.053 \pm 0.030$ | $0.055 \pm 0.032$ |
| 10°4 × 10°4 | $0.048 \pm 0.012$ | $0.051 \pm 0.014$ | $0.051 \pm 0.014$ | $0.048 \pm 0.017$ |
| | | 1°0 resolution | | |
| 5°2 × 5°2 | $0.046 \pm 0.027$ | $0.047 \pm 0.030$ | $0.049 \pm 0.037$ | $0.042 \pm 0.041$ |
| 10°4 × 10°4 | $0.049 \pm 0.015$ | $0.048 \pm 0.017$ | $0.049 \pm 0.018$ | $0.046 \pm 0.021$ |

[a]Table entries are the median and *rms* of 100 maximum likelihood $\Omega_b$ determinations for nominal values of the other parameters (see §4.).

[b]The modeled receiver noise is uncorrelated from pixel to pixel. Note that for a *fixed* observing duration, the noise per pixel in a 10°4 patch would be twice that in a 5°2 patch.

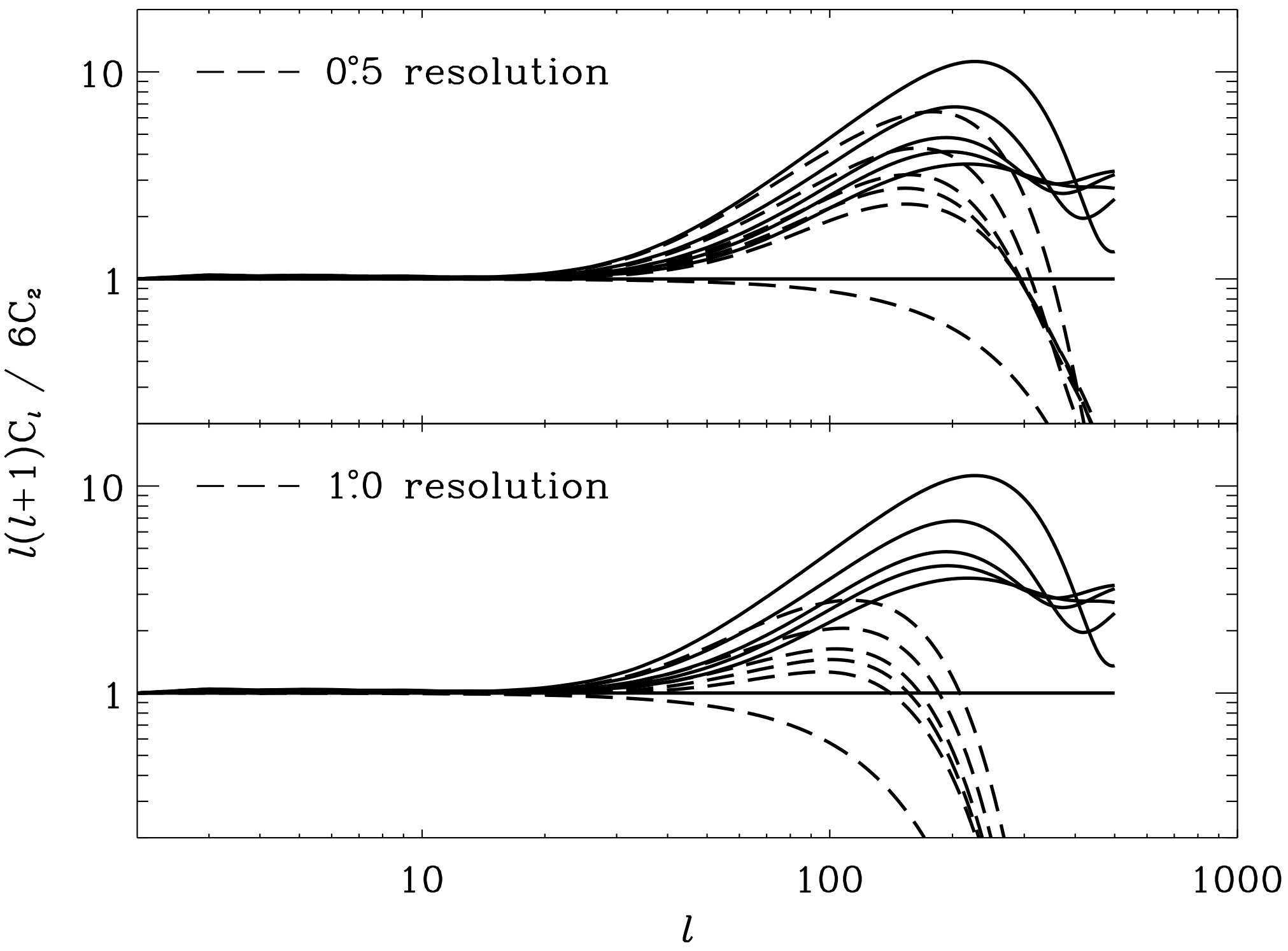

Fig. 1.— Spectra of the simulated CDM models, after Stompor (1994). All spectra have a scale-invariant primordial spectrum ($n = 1$), and have been normalized to a quadrupole anisotropy of $Q_{rms-PS} = 20$ $\mu$K. The flat model corresponds to a pure Harrison-Zel'dovich spectrum, and is included as representative of models without a Doppler peak. The CDM models plotted have a Hubble constant $H_0 = 50$ km s$^{-1}$ Mpc$^{-1}$, and baryon density $\Omega_b = 0.01, 0.03, 0.05, 0.10, 0.20$ increasing from bottom to top. The dashed lines in the top (bottom) panel correspond to spectra convolved with a 0°.5 (1°.0) FWHM Gaussian beam.

Color Plate

Color Plate

Color Plate

Fig. 2.— Simulated full-sky CMB maps at $0^\circ\!.5$ resolution. The temperature range, as indicated by the color bar, is $\pm 300\ \mu$K. *top*) Map generated with a flat spectrum model (see text). *middle*) Map generated using a CDM spectrum with baryon density $\Omega_b = 0.05$ and nominal values for the other parameters (see text). *bottom*) Same as middle, except $\Omega_b = 0.20$.

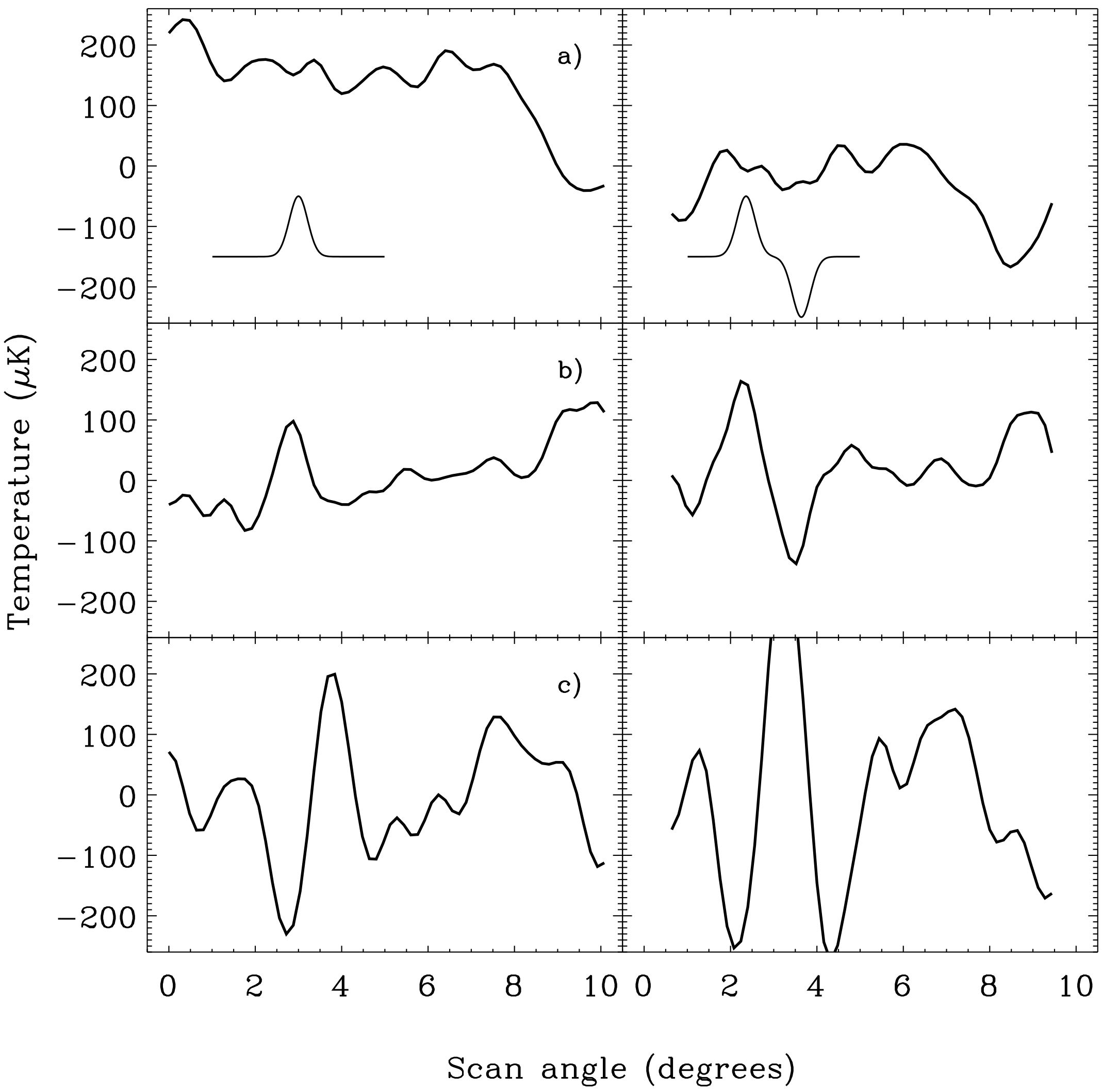

Fig. 3.— Selected map profiles (*left*) and corresponding single difference scans, with a 1°.28 beam throw (*right*). *a*) Profile from the flat spectrum model, and its corresponding single difference scan. *b*) Same as *a*) with an $\Omega_b = 0.05$ CDM model. *c*) Same as *a*) with an $\Omega_b = 0.20$ CDM model.

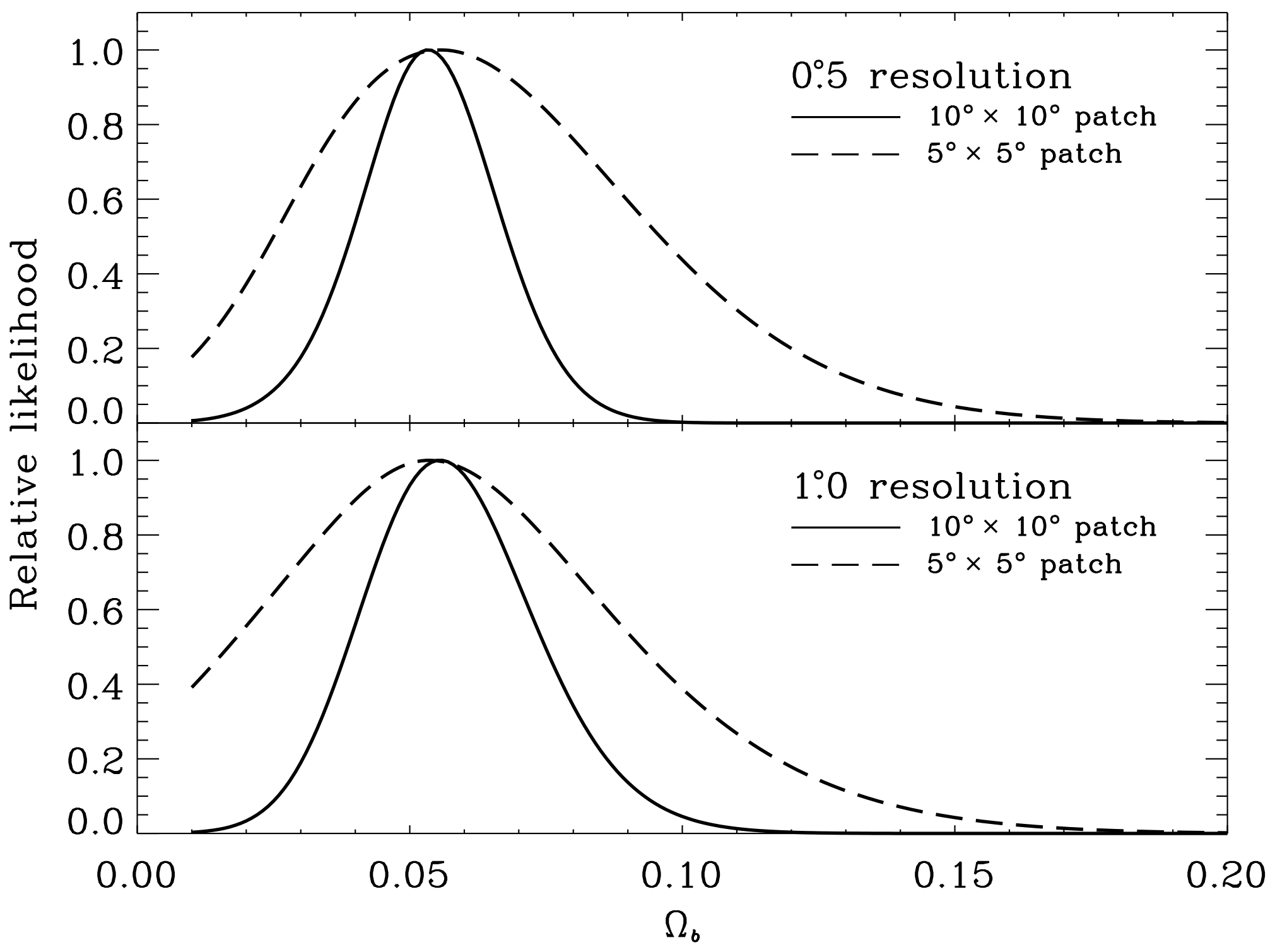

Fig. 4.— Likelihood functions for the baryon density $\Omega_b$ for selected sky patches. In each case the noise per pixel was 10 $\mu$K, and the input value of $\Omega_b$ was 0.05. Note that the width of the likelihood depends relatively strongly on the patch size, but not on the resolution.

---